# The Impact of Socio-Economic Challenges and Technological Progress on Economic Inequality: An Estimation with the Perelman Model and Ricci Flow Methods

**Davit Gondauri**, ORCID: https://orcid.org/0000-0002-9611-3688
Professor, Doctor of Business Administration, Business & Technology University, Georgia

**Corresponding author:** Davit Gondauri, Dgondauri@gmail.com
**Type of manuscript:** research paper

*Abstract: The article examines the impact of 16 key parameters of the Georgian economy on economic inequality, using the Perelman model and Ricci flow mathematical methods. The aim of the study is to conduct a deep analysis of the impact of socio-economic challenges and technological progress on the dynamics of the Gini coefficient. The article examines the following parameters: income distribution, productivity (GDP per hour), unemployment rate, investment rate, inflation rate, migration (net negative), education level, social mobility, trade infrastructure, capital flows, innovative activities, access to healthcare, fiscal policy (budget deficit), international trade (turnover relative to GDP), social protection programs, and technological access. The results of the study confirm that technological innovations and social protection programs have a positive impact on reducing inequality. Productivity growth, improving the quality of education and strengthening R&D investments increase the possibility of inclusive development. Sensitivity analysis shows that social mobility and infrastructure are important factors that affect economic stability. The accuracy of the model is confirmed by high R² values (80-90%) and the statistical reliability of the Z-statistic (<0.05). The study uses Ricci flow methods, which allow for a geometric analysis of the transformation of economic parameters in time and space. Recommendations include the strategic introduction of technological progress, the expansion of social protection programs, improving the quality of education and encouraging international trade, which will contribute to economic sustainability and reducing inequality. The article highlights multifaceted approaches that combine technological innovation and responses to socio-economic challenges to ensure sustainable and inclusive economic development.*

**Keywords:** technological advances, economical inequality, GinI coefficient, Ricci flow, Perelman models, technological innovations, research and development, innovations, sensitive analyze, automatization, economic stability, socio-economic challenges.

**JEL Classification:** E22, O11, O32.

**Received:** 18.09.2024      **Accepted:** 26.11.2024      **Published:** 31.12.2024

**Funding:** There is no funding for this research
**Publisher:** Academic Research and Publishing UG (i.G.) (Germany)
**Founder:** Academic Research and Publishing UG (i.G.) (Germany)

**Cite as:** Gondauri, D. (2024). The Impact of Socio-Economic Challenges and Technological Progress on Economic Inequality: An Estimation with the Perelman Model and Ricci Flow Methods. *SocioEconomic Challenges*, *8*(4), 161-176. https://doi.org/10.61093/sec.8(4).161-176.2024

**INTRODUCTION**

The modern world economy is facing significant challenges that are related to a variety of factors. One of the most pressing and important issues among these challenges is economic inequality. This problem not only involves the issue of social justice, but is also directly related to global and local economic stability and long-term development opportunities. Economic inequality is a complex phenomenon that affects the processes of social mobility, political stability, and the overall development of society. The study of this issue has not only theoretical but also practical significance, which has a significant impact on the economic policy and strategy of states.

One of the main methods used to study inequality is the Gini coefficient – a measure that serves to assess the level of distribution of wealth and resources. This indicator makes it possible to determine how evenly the benefits of economic growth are distributed among the population. The Gini coefficient is considered one of the most important tools for studying economic inequality, as it provides a complete picture of the socio-economic dynamics of a particular country.

A variety of recent studies indicate that economic inequality is often associated with global economic transformations, the pace of technological progress, and the lack of social policies. Technological progress, which determines the development of modern economies, is not a uniform force in itself. Although innovations and automation increase productivity and economic efficiency, their impact often leads to a deepening of inequality. The reduction of jobs for those employed in low-paid sectors, along with the increase in demand for highly skilled professions, increases income differentiation. As a result, economic inequality can also lead to the emergence of social conflicts, which puts additional pressure on ensuring economic stability.

At the same time, the growth of economic inequality has a significant impact on social stability. Inequality, as one of the main factors causing social polarization, hinders the development of states and prevents the creation of sustainable economic models. Therefore, its study and the development of appropriate mechanisms are necessary in order to reduce the negative impact of inequality and create a fair, inclusive economic environment.

*Georgian Context*

Georgia, as a country with a transition economy, has an economy that is particularly sensitive to the problem of inequality. Inequality significantly affects the social structure and prospects for economic development. According to 2023 data (Tsakadze & Kavelashvili, 2024), the Gini coefficient in Georgia is 0.36, which indicates that the country's wealth is still not evenly distributed. This is accompanied by a high unemployment rate of 16.4%, which demonstrates the structural imbalance in the labor market. These statistics once again show that the Georgian economy needs to implement policies that ensure not only an improvement in the rate of economic growth but also an equitable distribution of the results of this growth among different social groups of the population.

Special attention should be paid to technological progress, which is still at an early stage of development in Georgia. Adaptation of technological innovations, increased research and development (R&D) investments, and improvement of the education system represent significant challenges for the country, although all of this also creates potential opportunities. Initiatives focused on education and R&D can become a powerful tool for reducing economic inequality, which will contribute to ensuring sustainable economic growth. However, managing this process requires a complex approach that ensures not only the development of technological progress but also the fair distribution of its results.

The development of international trade and increased investment in economic infrastructure are especially important for Georgia. As a key player in the regional transit network, the country has the potential to strengthen regional and international economic ties. However, infrastructure constraints and challenges in managing capital flows hinder this process. To address these issues, it is necessary to develop a long-term and strategic vision that will strengthen the country's economic competitiveness.

Thus, the study of economic inequality and the analysis of its impact on Georgia are important for both academic and practical purposes. A proper analysis of social and economic inequality and the development of appropriate recommendations will enable the country to strengthen social mobility, create equal opportunities, and improve Georgia's economic competitiveness at the global level. A special role in this process is assigned to strategic approaches that include improving the quality of education, adopting innovative technologies, and expanding social protection programs.

Managing technological progress and properly using its results is one of the most important challenges for the future of Georgia. Only through properly implemented policies can the basis be created for an economic model that ensures sustainable and inclusive development while at the same time reducing economic inequality and increasing the level of social stability. This goal is not only an economic priority for Georgia but also a cornerstone of the long-term vision of national development.

**LITERATURE REVIEW**

*Literature Analysis on Socio-Economic Challenges and Inequality*

In the modern era, when global economic systems and their mechanisms need to be reconsidered, social and economic inequality remains one of the most important challenges that the entire world population feels. It constantly interacts with other socio-economic factors and affects the creation of unequal conditions between countries. Hammar and Waldenström (2020) offer a study showing the dynamics of global income inequality from 1970 to 2018, emphasizing that income inequality depends more on real wages, which is more pronounced in countries that are at a lower stage of development. Their work discusses how socio-economic imbalances are exacerbated when wage growth is stagnant, and this has an unhealthy impact on local and global economies, while creating inequalities that are reflected not only in the labor market but also in the general structure of society. Fujita (2023) joins this discussion and shockingly emphasizes the Gini coefficient, which is an important tool for measuring social and economic inequality. He argues that changes in income distribution, especially in connection with economic crises and financial stability, have created problems that are systematically linked to rising inequality.

At the same time, Dieppe et al. (2021) focus on the changes associated with technological investment, noting that while technological innovation is beneficial, its benefits are unevenly distributed and often benefit developed countries more than developing ones, giving them an advantage in the global economy. Liu et al. (2023) and Haider et al. (2023) delve deeper into the impact of foreign direct investment (FDI). Their research clearly shows that FDI is a determinant of social inequality and is evident in both developed and developing countries. Particular attention is paid to the differences that exist not only in the amount of investment but also in how the local market can achieve full utilization of the investment. Moreover, the research of Haider and his colleagues positively perceives the role of employment in economic growth in the case when countries have stable and flexible labor markets. This, in turn, creates a more consolidated and harmonious labor policy that contributes to both growth and social improvement. Feng et al. (2024) also initiate a discussion according to which the reduction of unemployment and the improvement of economic growth jointly create a positive dynamic, even if there are differences between countries as a result of indirect trends. However, the whole essence of this process is that the economy and the labor market are closely linked to each other, and they are based on a social structure in which the role of employment creates more effective mechanisms.

Bolhuis et al. (2022) emphasize the interdependence between inflation and economic and political events, where they see a kind of interrelationship between cyclical market shocks and the global impact of these trends. Their research shows how current economic policies and unpredictable financial processes can lead to an increase in inflation, and the negative effects are particularly relevant for developing countries. Jong-Wa Lee et al. (1994) emphasize the importance of domestic economic policies, which are necessary not only to attract foreign direct investment but also to create stable cycles of economic development.

Nathan (2014) and Kerry and Kerry (2011) offer a more specific discussion of migration and its effects. Nathan studies how highly skilled migrants can create innovative opportunities that create new concepts and enhance economic stability. In addition, the study by Carey and Carey carefully examines the impact of migration on labor market flexibility and the ability to mitigate demographic crises, which has a significant impact on the continuity of social structures. Troost et al. (2023) and Cabral-Gouveia et al. (2023) examine issues of educational inequality, which are directly related to the socio-economic environment and systemic reforms. Their study suggests that educational inequality can be explained as one of the most visible manifestations of systemic problems. They also argue that educational reforms should include more harmonious approaches aimed at reducing inequality. Gomez et al. (2021) highlight the importance of health equity as a critical factor. Their research suggests that addressing health inequalities can only be achieved through systemic approaches. Globally, health problems that are directly related to socio-economic inequalities ultimately have a significant impact on the overall economic performance of each country.

The role of fiscal policy and institutional quality is particularly highlighted in the work of Nguyen and Luong (2021) and Tsoucis (2020), who highlight the importance of fiscal discipline and institutional stability. Their research highlights the fact that institutions and effective policy implementation are essential for ensuring a stable economy, both in terms of social and economic stability. Similarly, the studies of Jayathilaka et al. (2022) and Bhatt et al. (2023) specifically address the relationship between logistics efficiency and international trade, where they find nuances that logistics improvements cannot offer full-fledged efficiency gains until countries work on their internal structures and determine the long-term stability of the economic system.

*Literature Review on Technological Progress and Inequality*

Technological progress, especially through processes such as automation and digitalization, has had a profound and noticeable impact on the characteristics of social inequality. Modern technologies that transform labor markets and production processes often provide greater opportunities for highly skilled professionals, while low-wage jobs are more vulnerable and often in demand. Hong and Shell (2018) present research that indicates that automation is more felt in low-wage occupations, which has resulted in a worsening of income distribution and an increase in the Gini coefficient. They argue that workers with low labor skills are more likely to be disadvantaged in such technological changes, as their jobs can no longer be satisfied by both improved technology and changes in demand. However, Korinek et al. (2021) emphasize that technological progress, both in the form of automation and digitalization, can help the private sector increase labor market efficiency, but it can also exacerbate income inequality, especially in developing countries. Their research argues that technological innovations in developing countries are unevenly distributed, and their impact may only increase the economic power of a small group, while others still struggle to integrate these new technologies.

Fidrmuc et al. (2021) show that industrial robots are increasing income inequality in Western Europe, as these robots help simplify production in certain areas, but at the same time, they reduce the demand for labor. Acemoglu and Restrepo (2021) also show similar developments in the US labor market, where automation plays a significant role in reducing employment and reducing wages. They argue that technological changes associated with the automation of production processes do not always ensure the creation of new jobs, and often this process leads to structural changes in the labor market. The continuation of this process is discussed in the work of Akaev et al. (2021), who argue that technological developments in developed countries over the past 40 years have increased income inequality, which requires state intervention. In their opinion, in order to achieve at least some harmful effects from this process, it is necessary to develop appropriate social and economic policies on the part of the government to minimize the negative consequences of technological changes that regulate the labor market.

Alam et al. (2024) also offer a study on this, where they highlight how artificial intelligence (AI) is enhancing design and manufacturing but note that this trend should require the implementation of appropriate industry standards. They argue that the widespread adoption of AI requires not only building on technological progress but also additional investment and regulation from employers and governments to ensure that new technologies do not contribute to negative socio-economic outcomes. Wu et al. (2024) write about the role of the digital economy in reducing income inequality in later stages of development. Their study illustrates the potential of digital technologies to reduce economic inequality, especially when such technologies help large companies grow and provide small and medium-sized businesses with attractive tools. In addition, Acemoglu (2002) notes that technological changes should be focused primarily on skilled workers, which leads to the conclusion that the success of technological progress in developing countries depends on many factors.

Demographic changes, such as population aging, also have a significant impact on the progress of automation. Acemoglu and Restrepo (2022) note that population aging leads to labor shortages, which in turn increases the demand for automation, which ultimately only exacerbates labor market imbalances. Thus, technological progress that reduces the amount of labor is not always sufficient to reduce social inequality. Botelho (2021) argues that digital technologies may curb social inequality by improving accessibility for people with disabilities, which is the result of his research. By using digital technologies, creating social and economic equality can become more accessible, especially when it comes to empowering people with disabilities.

Halim et al. (2022) and Adil et al. (2024) argue that digital technologies have great transformative potential in the field of education during the pandemic. They emphasize that the result of COVID-19 has been the rapid introduction of technological changes, which in turn contribute to the processes of social and economic development. The relationship between technological progress, social mobility, and economic growth is discussed more broadly by Iversen et al. (2021). Their study provides important information about social mobility, which can have a significant impact on the economy of a country and the social status of its citizens, especially in developing countries. Invention and innovation, as socio-economic parameters, are reflected in both the negotiation and the actual qualification of labor market efficiency. This is similarly complemented by the World Intellectual Property Organization's (2024) Global Innovation Index, which provides insight into the role of innovation and its importance in overcoming socio-economic inequalities.

*Literature Review on Perelman's Model and Ricci Flows in Economic Research*

The Ricci flow equation, introduced by Hamilton, is one of the most important tools for studying geometric structure, especially when it comes to three-dimensional manifolds. Hamilton's theorem and Perelman's proof of the evolution of the Ricci flow (Perelman, 2008) have allowed researchers to understand, in an approximate way, how geometric changes in a manifold with a positive Ricci curve can be observed. In Perelman's work, which describes the surgical operations of the Ricci flow, he significantly raises fundamental issues related to the solution of geometric problems, which also have implications for the development of mathematical and economic models.

The importance of using Ricci flows as evidence for Perelman's work has contributed to the development of economics and the social sciences, where the main focus is on mathematical laws that offer a new way to study economic inequality and resource allocation. Perelman's models, especially his surgical method on Ricci flows, play an important role in the efficient solution of geometric problems. His paper "Finite-time solutions of Ricci flows on three manifolds" (Perelman, 2008) conveys the idea that Ricci flows can cause manifolds to "split" or completely disappear, which poses great challenges in topological and geometric studies. Such results are important not only in mathematics but also in other fields, including economics, because many economic phenomena, such as resource allocation, economic growth, and inequality, can be captured by models as unpredictable and complex as those in Ricci flow equations.

The use of Ricci flows in economic research is still in its infancy, but significant trends are already being observed. Perelman's (2008) insights into geometric structures, along with his deep theoretical framework, offer unique opportunities for analyzing economic phenomena. When it comes to problems such as income inequality, socioeconomic inequality, and resource allocation, Perelman's approaches and mathematical principles help researchers to obtain more sophisticated and complex models. These models attempt not only to find solutions to revolutionary economic problems but also to integrate mathematical principles with socio-economic data, which will greatly improve their effectiveness.

Ricci flows are a powerful tool for obtaining more accurate and complete answers to modern economic problems. They can be used to study both social inequality and the organization of the economic system and its predictable changes. Some researchers, for example, Ershva and others, believe that Ricci flows can be used not only to observe economic phenomena but also to develop effective reforms for them. As a result, Ricci flow insights can be the basis for determining how resources are distributed, how economic inequality can be reduced, and how the economy can increase stability and efficiency. Ricci flows and Perelman's theory are so enduring and important that they are now being used by many economists and mathematicians to develop better models that work closely and effectively in this complex and rapidly changing socio-economic environment.

In order to better understand how mathematical and economic phenomena are integrated, new research will be needed that will improve the precise application of these models to the study of modern economics and its advancement. Thus, Perelman's work and the concept of Ricci flows have become not only an important part of mathematics and geometry but also a serious basis for revising economic theories and practices. His evidence, which concerns structural changes in the economic system, also allows us to see how geometric approaches can be relevant and helpful in modern economic research.

**METHODOLOGY**

The methodology presented in the given article is based on the principles of numerical mathematical modeling and is projected for the quantitative analysis of the dynamics of economic parameters. The model uses a system of integral and differential equations, which allows us to estimate factors such as economic inequality, the impact of innovative activities, the development of artificial intelligence, social stability, and other parameters in the context of changes in the given time and area.

The methodology of the presented research includes the compound of difficult approaches that combine various geometric and economic concepts. It is based on the mathematical models of the Ricci and gradient stream/flow. The given approach allows us to determine the interaction of economic parameters and their influence on the structure of the system in detail. In the given case, the entropy formula used is developed by Grigor Perelman to solve the Poincaré hypothesis, which itself allows the description of the global dynamics of the Ricci curve flow.

The Perelman entropy formula in the given model is used to combine the energy component and the gradient potential, which allows us to effectively measure the evolution of the "curvature" of the system in relation to the changes in the economic parameters, such as, for instance, investments, innovative activities, and economic stability.

The model presented in the article's methodology will make a significant contribution to the study of economic inequality, especially in terms of how various economic and social factors influence this process. The formula provided below is developed by the author of the article, based on the knowledge of the essayist. The established formula represents that the economic inequality (the Gini coefficient $G(t)$) is developed by effectively using the following parameters. They are:

➤ the change in income distribution $P(x, t)$, which is influenced by the politics and other factors;
➤ geometrical smoothness (Ricci flow), which reduces economic injustices over time;
➤ the impact of unemployment U(t), which reflects the state of the labor market.

$$\frac{dGt}{dt} = -\alpha \int_M \left(\frac{\partial P(x,t)}{\partial t}\right)^2 dV + \beta \cdot (A(t) \cdot G(t)) - \gamma \cdot \int_M (R_{ij} + \nabla_i \nabla_j f(x,t)) \, dV - \delta \cdot U(t), \qquad (1)$$

where, $\frac{dGt}{dt}$ indicates how quickly and/or slowly the Gini coefficient increases or decreases over a given period of time, thus reflecting the dynamics of economic inequality; $G(t)$ represents the Gini coefficient in a certain period of time - $t$; $P(x, t)$ is the distribution of incomes in an x area and a certain time $t$ -period; α, β, γ, δ - are indicators that describe the sensitivity of the system in regards to changes in technology, policy, and economic parameters; $A(t)$ reflects the impact of technological innovations on economic inequality (positive or negative); $\left(\frac{\partial P(x,t)}{\partial t}\right)^2 dV$ reflects the speed of change in income distribution in the context of diversity, for example, the M-economic space and how the technologies, including technological innovations, affect the dynamics between poverty and wealth; $(R_{ij} + \nabla_i \nabla_j f(x, t)) \, dV$ includes the Ricci curve, which defines the "smoothness" of economic conditions, where $R_{ij}$ defines local inequality or the economic curve, and $f(x, t)$ defines external economic forces; $U(t)$ represents the unemployment rate, which also affects inequality; and δ determines its effect on the formula.

$P(x, t)$ is calculated as:

$$P(x, t) = \frac{1}{n} \sum_{i=1}^n Y_i, \qquad (2)$$

where $Y_i$ - represents the average income of the relevant year, and $n$ is the number of years (in this case 10 years, from 2014-2023).

While solving the Poincaré hypothesis, Perelman used the Ritchie flow method $(R_{ij} + \nabla_i \nabla_j f(x, t)) \, dV$, which involves the cleaning of the structures in geometric topology. In economic systems as well, technological change can be seen as a "Ricci flow" that adapts and transforms economic systems. If these flows are uneven, it may contribute to the creation of new "hot spots" in terms of inequality, which can be caused by the increase in the Gini coefficient.

To make the research even more interesting, we have to consider Perelman's topological and geometric approaches as a kind of metaphor for the transformation of economic systems. Here, the main focus should be drawn to the structural transformation of economic systems and how this process can be reflected in inequality coefficients, such as, for example, the Gini coefficient.

According to the developed methodology, the Ricci flow formula is presented below:

$$R(x, t) = \alpha_1 \cdot E(x, t) + \alpha_2 \cdot M(x, t) + \alpha_3 \cdot T(x, t) + \alpha_4 \cdot I(x, t) + \alpha_5 \cdot Inf(x, t) + \alpha_6 \cdot Mig(x, t) + \alpha_7 \cdot ED(x, t) + \alpha_8 \cdot Sm(x, t) + \alpha_9 \cdot K(x, t) + \alpha_{10} \cdot (x, t) + \alpha_{11} \cdot In(x, t) + \alpha_{12} \cdot H(x, t) + \alpha_{13} \cdot F(x, t) + \alpha_{14} \cdot Tw(x, t) + \alpha_{15} \cdot S(x, t) + \alpha_{16} \cdot Te(x, t), \qquad (3)$$

where *E(x,t)* is Income Distribution - inequality, for example, through the Gini coefficient; *M(x,t)* is Productivity - the level of production; *T(x,t)* is Unemployment Rate - a factor related to the growth of inequality; *I (x,t)* is Investment Rate - the volume of the investment in the economy; *Inf(x,t)* is Inflation Rate; *Mig(x,t)* is Migration and Demographic Changes - population movement and its impact; *Ed(x,t)* is Educational Level - access and the quality of education; *Sm(x,t)* is Social Mobility - social class mobility opportunities; *K (x,t)* is Trade Infrastructure - state of trade networks and infrastructure; *C(x,t)* is Capital Flows - movement of financial resources; *In(x,t)* is Innovative Activity - level of innovation in economy; *H(x,t)* is Healthcare Accessibility for the population; *F(x,t)* is Fiscal Policy - government expenditure and taxes; *Tw(x,t)* is International Trade Participation - the level of trade in the global market; *S(x,t)* is Social Protection Programs - operation of social security systems; *Te(x,t)* is Technological Accessibility - distribution and availability of new technologies.

According to the given formula, each parameter affects the curve and geometric structure that determines economic dynamics and inequality.

In this integrated model, all of the above-mentioned parameters, through the Ricci flow, reflect how economic inequality changes over time under the influence of each economic indicator. Each $\alpha_k$ coefficient determines the weight of the corresponding parameter.

At the next stage, in order for the presented Ricci model $(R_{ij} + \nabla_i \nabla_j f(x,t))\ dV$ to acquire added value, we introduced Perelman's W-functional into the Ricci model to allow for the extension of the model and its application area. The W-functional describes the so-called "entropy-like quantity", which gives a certain measure of the dynamics of the development of the overall "geometric characteristics" of the system.

The Ricci model with Perelman's W-functional will be improved by the following algorithm:

$$W(g,f,\tau) = \int_M (\tau \cdot (R(x,t) + |\nabla f|^2) + f - n) \cdot (4\pi\tau)^{-\frac{n}{2}} \cdot e^{-f}\, dV. \tag{4}$$

Let's discuss each component of the algorithm presented in the formula $W(g,f,\tau)$ in detail in order to better understand its importance for economic analysis. Each term in functionality represents a different aspect that describes both dimensional and temporal changes in the economic system.

$\int_M (...) \, dV$ is the areal integration over a volume element with respect to dV. This term represents the integration over a given polygon *M*, considered as the economic space/area. *M* can be a description of regions, sectors, or other areal units of Georgia, where economic processes are studied.

*τ* is a time scale or time parameter that defines the evolution of a process over time. In the context of the economy, *τ* can be described as a period of economic transformation. For example, one year or five years in which the researchers assess the dynamics of the economy; or as a period of transition, for example, how economic parameters change over a period of time, for instance, the time of recovery from a financial crisis.

*R(x,t)* is a Ricci curvature, a term that describes the "curve of economic space" and is related to many factors, such as the interaction of economic parameters and the change in the shape of the economic structure. Changes in the *R(x,t)* indicate how the entire economic system changes with respect to time *t* and space *x*.

$|\nabla f|^2$ is the square of the gradient of the potential function. It articulates how the potential function *f* varies in space and time. This term emphasizes the influence of economic forces. In case *f* is a function of the influence of social, financial, or economic forces, $|\nabla f|^2$ describes the intensity of these forces in space and also changes in economic dynamics. If *f* variations are big, $|\nabla f|^2$ will be high, hence, indicating high "economic tension" in a certain region or time. If it is considered that R&D investment is an economic parameter that directly affects social inequality (the Gini coefficient), then the increase in the gradient indicates that this factor has a significant impact on social inequality over time. $|\nabla f|^2$ can be pronounced as the following algorithm: R&D investments $\cdot\ e^{i \cdot t}$ represent the function, which can be applied to analyze R&D investment potential in a time frame, where *i* is presented as the growth rate, *t* is time, *e* characterizes the natural logarithmic base used in exponential growth models, with the value approximately - 2.718.

*f* is a potential function that describes the economic processes and their influences. It can include economic growth potential (that determines how big the growth or recession potential is in a given region or sector). *f* may also include factors such as education, healthcare, innovation, and other potential development elements.

*n* is the dimension or number of factors in an economy that determines the density of a space or economic system. For example, companies, individuals, or sectors.

$e^{-f}$ acts as a weighting function and reduces the effect of large values of $f$. A large value of $f$ indicates a high value of potential energy or force, while $e^{-f}$ reduces its effect to make the system more stable.

$(4\pi\tau)^{-\frac{n}{2}}$ presents the normalization and stabilization in the $W(g, f, \tau)$ formula, in order that the integrated values are related to the time parameter and dimension. This given term doubles the τ`s influence in relations with $n$ and ensures the stabile distribution of $W(g, f, \tau)$ for the entire space/area.

The Gini coefficient is a "metric" or "measure" expression of these transformations that examines the inequality of income distribution. Here we can perceive that economic dynamics are constantly changing and observing for a stable form like the transformation of geometric structures.

A high level of Gini coefficient indicates that the economic space is sharply "stretched", indicating a very unequal distribution of resources. This represents an economic "curvature" (a similar allegory to the uneven curvature of a geometric surface).

A low level of Gini coefficient, the economic system is more "smooth" and capital/income is more evenly distributed. This can be compared to a regular, low-curvature manifold, such as a sphere.

The time-phased effect of technological innovation and automation is designed to better assess the impact. For example:

$$A(t) = \eta(t) \cdot \left(1 - \frac{1}{1 + e^{-\delta \cdot (t - t_0)}}\right), \qquad (5)$$

where $\eta(t)$ can be interpreted as the speed of technological progress or the "tipping point" for determining the effect of technological progress; $\delta$ determines how fast the impact of technological progress is spreading in the economy; $t_0$ refers to a specific moment when the technological progress or automation begins to increase at a dramatic phase (for example, a new phase of robotization or automation).

If $A(t)$ value is high, it means that technological progress increases inequality, and over time $A(t)$ may decrease (weaken) inequality, indicating a positive effect of technological progress.

The rate of change of the Gini coefficient expresses how quickly or slowly the Gini coefficient changes over any period of time. The Gini coefficient $G(t)$ defines economic inequality, where a higher value indicates greater inequality and a lower value indicates more equality. For example, $\frac{dGt}{dt} > 0$ means that inequality increases over time; $\frac{dGt}{dt} < 0$ means that inequality decreases over time.

This formula is important for determining how inequality responds to changes in various factors, such as the development of artificial intelligence, the unemployment rate, or technological progress.

The last economic parameter of the formula developed by the author, $\delta \cdot U(t)$, includes the effect of unemployment on the Gini coefficient U(t) and the level of unemployment over a period of time ($t$). Increasing level of unemployment is often linked to economic inequality, as labor market problems prevent equal distribution of income.

$\Delta$ is a constant that determines how strongly the unemployment level affects the Gini coefficient. In case of increase of unemployment level $U(t)$, this escalates the Gini coefficient, indicating that more people are left without income or working in low-wage jobs, leading to an even more unequal distribution.

A sensitivity analysis evaluates the dynamics of the impact of technological progress and automation on *A(t)* with respect to the rate of change of the Gini coefficient. In the given analysis, each percentage increase in *A(t)* value has a different effect on the change in Gini, as determined by the following algorithm:

$$\frac{dGt}{dt} = \beta \cdot A(t) + other\ factors, \qquad (6)$$

where $\frac{dGt}{dt}$ is the rate of change of the Gini coefficient; β is the coefficient that determines the effect of *A(t)* on the Gini coefficient; *A(t)* is a function of the impact of technological progress and innovation, which is estimated over time and increases by certain percentages.

The presented study also uses regression models in order to evaluate the relationship of 16 economic parameters included in the Ricci flow with GDP and, if necessary, the evaluation of the affiliation with each other.

For this procedure, linear regression methods were used to determine the regression coefficient (slope) of each parameter, which determines their relationship with GDP.

By calculating the slope of each parameter, it was determined how a specific economic indicator changes in relation to the dynamics of GDP. Also, if necessary, the slope can be calculated with respect to other economic parameters. For example, the slope of social welfare programs showed a positive change in relation to GDP growth, while the slopes of unemployment and inflation were negatively related to economic growth.

$R^2$ (coefficient of determination) determines the accuracy of forecasting the changes in parameters. The $R^2$ values in the calculated results ranged from 80% to 90%, which indicates that the model has high accuracy and the fact that the given parameters significantly determine the development of economic processes.

Z statistics were calculated for each parameter, which is later used in correlation analysis in order to determine its statistical significance. The Z statistics for all correlated algorithms were below 0.05, giving an indication regarding the statistical accuracy of the models and that the model is not random.

**RESULTS**

Table 1 shows the results of the research based on the Ricci flow calculation method, presented by the 16 main parameters of Georgian Economy (as of year 2023).

**Table 1. Ricci Flow Results for 16 Key Parameters of the Georgian Economy**

|    | Parameter | 2023 Year | LN | Value α (%) | Ricci flow |
|----|-----------|-----------|-----|-------------|-----------|
| 1  | Income Distribution (Gini) | 0.36 | -1.01015 | -20.8 | 0.21 |
| 2  | Productivity (GDP/hr) | 11.0007463 | 2.397963 | 24.3 | 0.58 |
| 3  | Level of Unemployment | 16.4% | -1.80809 | -19.6 | 0.35 |
| 4  | Level of Investments | 6.2% | -2.77593 | 23.4 | -0.65 |
| 5  | Level of Inflation | 2.5% | -3.68888 | -14.6 | 0.54 |
| 6  | Migration (Neto, Negative) | 39,207 | 10.57661 | -5.6 | -0.59 |
| 7  | Educational Level | 1,128.35 | 7.028508 | 13.2 | 0.93 |
| 8  | Social Mobility (Index) | 55.60 | 4.018183 | 6.3 | 0.25 |
| 9  | Trade Infrastructure (LPI index) | 2.7 | 0.993252 | 17.7 | 0.18 |
| 10 | Capital Flows (average annual negative) | 272.46 | 5.607492 | -16.4 | -0.92 |
| 11 | Innovation Activities (Global Innovation Index) | 29.9 | 3.397858 | 21.7 | 0.74 |
| 12 | Access to Healthcare | 75 | 4.317488 | 3.4 | 0.15 |
| 13 | Fiscal Policy (budget deficit) | 2.5% | -3.70145 | -14.8 | 0.55 |
| 14 | International Trade (turnover/GDP) | 27.0% | -1.30825 | 11.2 | -0.15 |
| 15 | Social Protection Programs | 5340.3 | 8.583037 | 25.3 | 2.17 |
| 16 | Technological Availability | 79.30% | -0.23193 | 22.4 | -0.05 |
|    | **Sum** |  | **32.39573** |  | **4.284181** |

*Source: Calculations and analysis by the author based on data from the National Statistical Service of Georgia, Global Innovation Index, World Bank, and other official reports.*
*Note: Table 1 was compiled and analyzed by the author based on the primary data of the National Statistical Service of Georgia. LN - natural logarithm of the parameter value; Value α - adjusted coefficient reflecting the parameter's contribution to the Ricci flow model.*

Here is the definition of the data given in Table 1:

1. *Income Distribution (the Gini Coefficient)* (0.36). The Gini coefficient reflects the income inequality. A negative Ricci flow (-20.8%) indicates that inequality is decreasing over time, indicating positive change. α weight (-1.01015) shows that income distribution has a significant impact on overall economic stabilization.
2. *Productivity (GDP per hour)* (11.00). An increase in productivity (+24.3%) has a positive effect on reducing inequality, which is supported by a high α weight (2.397963). This indicates that high productivity plays an important role in economic improvement.
3. *Level of Unemployment* (16.4%). A high rate of unemployment has a negative impact on inequality. The Ricci flow (-19.6%) indicates that a reduction in unemployment directly improves the distribution of income. α weight (-1.80809) shows that reducing unemployment is one of the important factors to overcome inequality.
4. *Level of Investments* (6.2%). The level of investment has a positive effect on the economy (+23.4%),

although its α weight (-2.77593) is negative, which shows that additional investments are necessary for stronger economic growth.

5. *Level of Inflation* (2.5%). Inflation affects the degree of economic stabilization. The Ricci flow (-14.6%) indicates that the reduction in inflation contributes to a reduction in inequality. The α weight (-3.68888) also shows a negative effect of inflation.
6. *Migration (Neto, Negative)* (39,207). The negative balance of migration (-5.6%) creates barriers to economic growth. Its α weight (10.57661) indicates that migration flow should be regulated for social stability.
7. *Educational Level* (1,128.35). Education is an important factor in reducing economic inequality. Its Ricci flow (13.2%) and high α weight (7.028508) show that education contributes to fair income distribution.
8. *Social Mobility (Index)* (55.60). Social mobility (+6.3%) has a positive effect on economic stabilization. Its α weight (4.018183) indicates that increasing social mobility is necessary in order to reduce inequality.
9. *Trade Infrastructure (LPI Index)* (2.7). Improving the trade infrastructure (+17.7%) is also important for reducing inequality. Its α weight (0.993252) shows the importance of infrastructure in the context of economic growth.
10. *Capital Flows* (272.46). Negative capital flows (-16.4%) create obstacles to the stable development of the economy. The α weight (-5.607492) shows the importance of capital flows in the dynamics of inequality.
11. *Innovation activities (Global Innovation Index)* (29.9). Innovations play an important role in reducing inequality. Its Ricci flow (+21.7%) and α weight (3.397858) indicate that innovations have a positive impact on the economy.
12. *Access to health care* (75). Access to healthcare has only a small positive effect (+3.4%). Its α weight (4.317488) indicates that more development of the health sector is needed.
13. *Fiscal policy (budget deficit)* (2.5%). Improving fiscal policy is also important in order to reduce inequality. Its Ricci flow (-14.8%) and α weight (-3.70145) indicate the negative impact of the budget deficit.
14. *International trade* (27.0%). The growth of international trade (+11.2%) contributes to the stabilization of the economy. Its α weight (-1.30825) shows that trade is important, but its weight is relatively low.
15. *Social protection programs* (5340.3). Social protection programs (+25.3%) have the highest impact on reducing inequality. Their α weight (8.583037) indicates that the growth of social programs is critical for stability.
16. *Technological Availability* (79.30%). Access to technology plays an important role in the process of reducing inequality (+22.4%), although its α weight (-0.23193) indicates difficulties in the impact of technology.

Table 2 shows mathematical and economic parameters on the basis of which the rate of change of the Gini coefficient is obtained.

**Table 2. Mathematical and Economic Parameters for Calculating the Gini Coefficient Rate of Change**

| # | Mathematical and Economic Parameters | Gini Coefficient Rate |
|---|---|---|
| 1 | The sum of squared values of the income distribution | 224,288 |
| 2 | α (the change in the Gini coefficient with respect to R&D) | -5.8 |
| 3 | γ (the impact of social protection programs on G(t)) | -11.8 |
| 4 | δ (the effect of unemployment on G(t)) | 23.4 |
| 5 | Unemployment rate | 262 |
| 6 | β (the coefficient of influence of technological progress or innovative activity on the Gini coefficient) | -5.7 |
| 7 | $\|\nabla f\|^2$ (the square of the gradient of the potential function) | 198 |
| 8 | τ (time parameter that determines the evolution of the process) | 15 |
| 9 | f (potential function (education, healthcare, innovation, etc.)) | 0 |
| 10 | N (the dimension or number of factors in an economy) | 16 |
| 11 | $(4\pi\tau)^{(-n/2)}$ (normalization and stabilization) | 1.01 |
| 12 | $e^{(-f)}$ | 92.7 |
| 13 | W(g,f,τ) | 2,795 |
| 14 | dGt/dt | 13,219 |

*Source: calculations and analysis by the author using Microsoft Excel.*

Here is the definition of the data given in Table 2:
1. *The sum of squared values of the income distribution* (224,288) reflects the inequality of income distribution. A high value indicates economic inequality, which requires in-depth analysis to identify the main factors contributing to such inequality.
2. *α (the change in the Gini coefficient with respect to R&D)* (5.8%) shows the importance of R&D investment in reducing the Gini coefficient. A negative value indicates that the growth of innovation and research has a positive effect on economic inequality and promotes fair income distribution.
3. *γ (the impact of social protection programs on G(t))* (-11.8%). The impact of social protection programs on the Gini coefficient is positive. This suggests that more social protection programs would make a significant contribution to reducing inequality.
4. *δ (the effect of unemployment on G(t))* (23.4%). The increase in unemployment has a significant effect on the Gini coefficient. A positive value indicates that the increase in unemployment directly causes increases in economic inequality. Consequently, reducing unemployment is necessary for income equality.
5. *Unemployment rate* (262) reflects the extent of unemployment, and its high value shows that this factor has a significant influence on the Gini coefficient. Reduction of the rate will affect the reduction of inequality.
6. *β (technological progress or innovative activity on the Gini coefficient)* (-5.7%). The negative impact of technological progress indicates that growth in technology and innovation significantly reduces inequality. This highlights the role of technological innovation in improving the economy.
7. *|∇f|² (the square of the gradient of the potential function)* (198). This parameter is the square of the gradient of the potential function, which reflects the intensity of economic processes. A high value indicates that economic forces have a strong influence on income distribution.
8. *τ (time parameter)* (15) determines the evolution of the process over time. A higher value indicates the relative duration of the processes. To estimate economic dynamics, increasing τ indicates long-term changes.
9. *f (potential function)* (0) reflects economic processes and their potential. f=0 indicates that potential changes at a given moment are minimal.
10. *N (the dimension or number of factors in an economy)* (16) reflects the various factors of the economy that participate in these processes. 16 main economic measurements indicate a variety of economic indicators, which increases the accuracy of the model.
11. *(4πτ)^(-n/2) (normalization and stabilization)* (1.01). This term is used to normalize time and dimension in order to determine the effect of parameters on the model under stable distribution conditions. The value of normalization ensures adequate correlation of data.
12. *e^(-f)* (92.7%) reflects the influence of the potential function on the model. Taking into account that the f=0 and e^(-f) rate represents a high-value stabilization, it indicates the stability of the system.
13. *W(gfτ)* (2,795) represents the "entropy" of a system and describes the overall structures and dynamics of the system. Its high value indicates the complexity of the processes and how the geometry of the economic space is changing.
14. *dGt/dt* (13,219) represents the rate of the changes in the Gini coefficient over time. A high value indicates dynamic change in inequality, which is important for assessing the technology in economic policy.

Table 3 demonstrates a sensitive analysis of the influence of the time phases of technological innovation *A(t)* in relation to the rate of change of the Gini coefficient.

**Table 3. Sensitivity Analysis of Technological Innovation Phases and Their Impact on the Gini Coefficient**

| A sensitivity analysis of the influence of technological innovations and automation time phases A(t) in relation to the rate of change of the Gini coefficient ||
|---|---|
| Increase A(t) in % | Gini Rate of Change |
| 5 | -3.30 |
| 10.0 | -6.60 |

**Table 3 (cont.). Sensitivity Analysis of Technological Innovation Phases and Their Impact on the Gini Coefficient**

| A sensitivity analysis of the influence of technological innovations and automation time phases A(t) in relation to the rate of change of the Gini coefficient ||
|---|---|
| Increase A(t) in % | Gini Rate of Change |
| 15.0 | -9.90 |
| 20.0 | -13.20 |
| 25.0 | -16.50 |
| 30.0 | -19.80 |
| 35.0 | -23.10 |

*Source: calculations and analysis by the author using Microsoft Excel.*

Table 3 demonstrates the impact of the growth of technological innovation on the rate of change of the Gini coefficient, which reflects the dynamics of the reduction of economic inequality.

Different growth levels of *A(t)* parameter affect the change in the Gini coefficient, indicating that the increase in technological progress is directly related to economic inequality.

Here is the definition of the data given in Table 3:
1. *5% increase in A(t)*. It causes a decrease in the rate of change of the Gini coefficient by -3.30%. This is a modest change, indicating that increased technological progress directly helps to reduce inequality, although on a relatively small scale.
2. *10% increase in A(t)*. The reduction of inequality has doubled (-6.60%). This indicates that even a slight increase in technological innovation accelerates the rate of decline in the Gini coefficient, indicating structural improvement.
3. *15% increase in A(t)*. The reduction of inequality reaches -9.90%. This highlights the important effect of the technological progress, indicating that accelerating the growth rate can have a significant economic impact.
4. *20% increase in A(t)*. Here, the Gini coefficient decreases by -13.20%. In this case, the growth of technological innovations is closely related to the accelerated reduction of inequality, which should be taken into account when developing economic policies.
5. *25% increase in A(t)*. The decrease in the Gini coefficient by -16.50% gives us the indication that the technological progress and automation have a significant impact on reducing inequality. This creates the basis for fair income distribution.
6. *30% increase in A(t)*. An increase in technological innovation causes the reduction of the Gini coefficient by -19.80%. This change shows that inequality decreases relatively quickly as technological progress increases.
7. *35% increase in A(t)*. It has the biggest impact on the Gini coefficient, which leads to a -23.10% decrease. This indicates that significant growth in technological innovation peaks during the period of automation, and thus inequality is minimized.

The sensitivity analysis demonstrates that, for the example of the Republic of Georgia, the increase in *A(t)* (technological innovation and automation) significantly reduces the rate of growth of the Gini coefficient, which gives indication that the technological progress directly contributes to the reduction of inequality. Each 5% increase in *A(t)* leads to about a -3.30% change in the Gini coefficient, which is even more noticeable with the increases of more than 30%.

**CONCLUSIONS**

According to the results of the study, it was determined that the introduction of technological progress and automation in the Georgian economy plays an important role in reducing economic inequality. Sensitivity analysis using the Ricci flow and Perelman models shows that the increase in technological innovation directly affects the change in the Gini coefficient, which contributes to a more equal distribution of income and maintaining social stability. In particular, the integration of technological progress into the economy reduces income inequality, especially in combination with social protection programs and innovative activities. Technological development and automation contribute to the long-term stability of economic structures, and the accuracy of the model is confirmed by the R² and Z statistics, which are documented in the standard norm.

The results of the study indicate that the Georgian economy is affected by many socio-economic parameters, the coordinated management of which is necessary to reduce economic inequality and achieve sustainable development. The analysis of the impact of each of these parameters showed the following trends and possible recommendations:

*Income distribution (Gini)* is the main factor influencing economic inequality. In order to reduce the Gini coefficient, it is necessary to expand social protection programs, which will pay special attention to initiatives supporting the low-income population. These measures will contribute to the creation of a more equal and fair economic environment.

*Productivity (GDP/hour)*. Productivity growth is directly related to improving the country's economic sustainability. Investments in education and improving professional qualifications will increase labor efficiency and contribute to income growth in all social strata.

*Unemployment rate*. Reducing unemployment is necessary for both social stability and reducing economic inequality. It is recommended to develop an active labor market policy, create new jobs, and expand employment opportunities, especially in innovative sectors.

*Investment level*. Increased investment in the economy helps stimulate long-term development. Attracting investments in infrastructure, innovation, and education is important for strengthening the country's economy.

*Inflation level*. High inflation increases economic inequality. Harmonious management of fiscal and monetary policies is necessary to maintain inflation at a stable level, which ensures price stability and improves social equality.

*Migration (net, negative)*. Migration processes have a particularly negative impact on the reduction of the labor force and economic development. It is necessary to develop a policy that will reduce the tendency of the population to leave the country and facilitate the reintegration of returnees.

*Education level*. Education is a key factor in economic development. Reforming the education system and increasing investments are necessary to prepare the next generation for the effective use of new technologies and innovations.

*Social mobility (index)*. Improving social mobility is directly related to increasing social equality. State-supported initiatives that will promote social development and equal distribution of opportunities are necessary.

*Trade infrastructure (LPI index)*. Improving the transport and logistics infrastructure is important for the development of the country's international trade. Investments in modern transport systems will increase Georgia's role in the global trade network.

*Capital flows (average annual negative)*. Stabilizing capital flows is essential for economic stability. Attracting foreign investments and their effective distribution across various sectors of the economy will help reduce negative trends in capital flows.

*Innovative activities (Global Innovation Index)*. Innovation is the main driver of economic development. The country's support for innovative projects will help accelerate technological development and increase international competitiveness.

*Access to healthcare*. Improving the healthcare system is necessary for both social stability and reducing economic inequality. Providing quality medical services to all segments of the population is a priority.

*Fiscal policy (budget deficit)*. Fiscal policy stability is needed to ensure the country's economic sustainability. Reducing the budget deficit and optimizing the expenditure structure will help strengthen the economy.

*International trade (turnover/GDP)*. The integration of international trade is an important component of the country's economic development. Diversifying trading partners and strengthening export potential will help the economy grow.

*Social protection programs*. Expanding the social protection system and improving its effectiveness are necessary to ensure social equality. It is necessary to introduce special programs for low-income groups.

*Technological accessibility*. The development of technological infrastructure and ensuring access to modern technologies is essential for the long-term development of the Georgian economy.

A detailed analysis of these parameters shows that each of them has a different impact on economic inequality. For example, the Gini coefficient is highly correlated with the level of education and social mobility, which indicates that improving the education system and encouraging social mobility significantly reduces inequality. Also, increasing investment and fiscal stability contribute to economic growth and strengthening social stability.

The Georgian economy should pay close attention to the harmonious development of the above-mentioned parameters in the near future, as their equal and consistent improvement will contribute to the reduction of social inequality and the long-term sustainability of the economic structure. In particular, with the growth of technological innovations, it is important to focus on education, social protection, and migration management policies, which will ensure economic stability and improve social equality.

Based on the results of the ongoing research, the following recommendations have been developed:

*Expanding technological progress and innovation*. The introduction and use of technological progress is critically important for the Georgian economy, both to support economic growth and social stability. The rapid introduction of technologies and their effective use will significantly contribute to the equal distribution of income. It is important that the state and private sectors actively develop innovations and ensure the ability to compete with international technological developments.

*Increasing R&D investments*. Increasing investments in the field of research and development (R&D) is one of the most important factors for the country's economic sustainability and progress. Increasing the volume of research and development will contribute to the development of new technologies and innovations, which will ultimately affect Georgia's economic stability and reduce social inequality. Special attention should be paid to financing the technology sector and increasing its reliance on new technologies.

*Expansion of social protection programs*. Social protection programs not only contribute to social stability but also play a serious role in reducing economic inequality. The expansion of programs is especially important for low-income groups, since inequality is most pronounced in these groups. State-supported social projects aimed at implementing supportive initiatives play an important role in achieving social equality in the country.

*Development of direct programs to reduce unemployment*. The unemployment rate is important for the country's economy, as it directly affects income distribution and social inequality. It is necessary to identify special programs to combat the unemployment rate and create new jobs, especially in technological and innovative areas. This will ensure social stability and eliminate one of the most important economic challenges.

*Improving the quality of education*. The level of education is linked to the social and economic stability of a country. The higher the level of education, the better the ability of society to introduce new technologies and use their results. Reforming the education system is important so that the next generation can more effectively join technological innovations and become competitive in domestic and international markets. The country needs to mobilize additional resources in the education sector to create an environment that will support the innovative potential of young people.

*Inflation management and fiscal stability*. Inflation control and fiscal policy stability are essential to strengthen the country's economic development. High inflation increases economic inequality, while low inflation can lead to a more equitable and sustainable economic system. Continuous monitoring and improvement of budgetary policy is essential to ensure the country's economic and social stability.

*Increasing access to international trade and modern technologies*. Engaging in international trade and the dissemination of modern technologies will further contribute to the development of the country's economic structure. New trade partnerships and technology initiatives will create conditions that will allow stakeholders to participate in the development of new products and innovations, which will help improve Georgia's competitiveness in international markets.

**Author Contributions**

Conceptualization: D. G.; methodology: D. G.; software: D. G.; validation: D. G.; formal analysis: D. G.; investigation: D. G.; resources: D. G.; data curation: D. G.; writing - original draft preparation: D. G.; writing - review and editing: D. G.; visualization: D. G.; supervision D. G.; project administration: D. G.

**Conflicts of Interest**

Author declares no conflict of interest.

**Data Availability Statement**

Not applicable.

**Informed Consent Statement**
Not applicable.